\documentclass[groupedaddress,aps,pra,superscriptaddress,showpacs,twocolumn,prl]{revtex4}%
\usepackage{epsfig,dsfont,amssymb,amsmath,amsthm,amsfonts,amsbsy,mathrsfs}
\usepackage{graphicx, color}
\usepackage{epstopdf}
\usepackage{mathdots}
\usepackage{float}
\begin{document}

\title{Generalized entanglement monogamy and polygamy relations for $N$-qubit systems}

\author{Zhi-Xiang Jin}
\thanks{Corresponding author: jzxjinzhixiang@126.com}
\affiliation{School of Mathematical Sciences,  Capital Normal University,  Beijing 100048,  China}
\author{Shao-Ming Fei}
\thanks{Corresponding author: feishm@mail.cnu.edu.cn}
\affiliation{School of Mathematical Sciences,  Capital Normal University,  Beijing 100048,  China}
\affiliation{Max-Planck-Institute for Mathematics in the Sciences, Leipzig 04103, Germany}
\author{Xianqing Li-Jost}
\affiliation{Max-Planck-Institute for Mathematics in the Sciences, Leipzig 04103, Germany}
\affiliation{School of Mathematics and Statistics, Hainan Normal University, Haikou 571158, China}
\bigskip

\begin{abstract}

We investigate the generalized monogamy and polygamy relations $N$-qubit systems.
We give a general upper bound of the $\alpha$th ($0\leq\alpha\leq2$) power of concurrence for $N$-qubit states. The monogamy relations satisfied by the $\alpha$th ($0\leq\alpha\leq2$) power of concurrence are presented for $N$-qubit pure states under the partition $AB$ and $C_1 . . . C_{N-2}$, as well as under the partition $ABC_1$ and $C_2\cdots C_{N-2}$. These inequalities give rise to the restrictions on entanglement distribution and the trade off of entanglement among the subsystems. Similar results are also derived for negativity.

\end{abstract}

\maketitle

\section{INTRODUCTION}
Quantum entanglement \cite{FMA,KSS,HPB,HPBB,JIV,CYS} as a physical resource in quantum communication and quantum information processing has been the subject of many studies in recent years. The study of quantum entanglement from different points of view has been a very active area and has led to many interesting results. Although bipartite entanglement is relatively well understood at least for qubit systems, the characterization or classification of multipartite entanglement is still very challenging.

Monogamy is very important in multipartite systems \cite{VJW,MP,AV}.  As one of the fundamental differences between quantum entanglement and classical correlations, a key property of entanglement is that a quantum system entangled with one of other systems limits its entanglement with the remaining ones. The monogamy relations give rise to the restrictions on distribution of entanglement in the multipartite setting. 
The monogamy relations was first noted by Coffman, Kundu, and Wootters \cite{VJW} based on an inequality satisfied by the squared concurrence $C$, often referred to as the CKW inequality. It is known to us concurrence is a proper measure satisfied the monogamy inequality introduced by Wootters \cite{SW,WKW}.

For a tripartite system $A$, $B$ and $C$, the usual monogamy of an entanglement measure $\mathcal{E}$ implies that the entanglement between $A$ and $BC$ satisfies $\mathcal{E}_{A|BC}\geq \mathcal{E}_{AB} +\mathcal{E}_{AC}$. For any three-qubit state $|\psi\rangle_{ABC}$, the CKW is
$C^2(\rho_{A|BC})\geq C^2(\rho_{AB})+C^2(\rho_{AC})$,
where $\rho_{A|BC}$ stands for the bipartite splitting between $A$ and $BC$.
The concurrence of a two-qubit mixed state $\rho$ is given by $C(\rho) = \mathrm{max}\{0,~\lambda_1-\lambda_2 -\lambda_3 -\lambda_4\}$, where $\lambda_1,~\lambda_2,~\lambda_3,~\lambda_4$ are the square root of the eigenvalues of $\rho(\sigma_y \otimes \sigma_y )\rho^{\star}(\sigma_y \otimes \sigma_y )$ in nonincreasing order, $\sigma_y$ is the Pauli matrix, and $\rho^{\star}$ is the complex conjugate of $\rho$. $\rho_{AB}$ and $\rho_{AC}$ are the reduced density matrices of $|\psi\rangle_{ABC}\langle\psi|$, respectively. Monogamy relations are not always satisfied by any forms of entanglement measures. The concurrence and entanglement of formation do not satisfy the usual monogamy inequality. However, the $\alpha$th ($\alpha\geq2$) power of concurrence and the $\alpha$th ($\alpha\geq\sqrt{2}$) power of entanglement of formation do satisfy the monogamy relations for $N$-qubit states \cite{XS}. In \cite{JF} a tighter monogamy relation for $\alpha$th ($\alpha\geq2$) of concurrence has been presented. Recently, in \cite{gy1,gy2}, the authors introduced a definition of monogamy and polygamy relations without inequalities.

In this paper, we
study the general monogamy inequalities with $AB$ as the focus qubits, satisfied by the concurrence and the concurrence of assistance (COA) \cite{CH}.
We give an upper bound for the $\alpha$th ($0\leq\alpha\leq2$) power of concurrence for multiqubit states. Then the monogamy relations of the $\alpha$th ($0\leq\alpha\leq2$) power of concurrence in $N$-qubit pure states under the partition $AB$ and $C_1 . . . C_{N-2}$, as well as under the partition $ABC_1$ and $C_2\cdots C_{N-2}$, are established. Based on the relation between negativity and concurrence, we also obtain the similar results for negativity.

\section{Improved generalized monogamy and polygamy relations of concurrence}

For a bipartite pure state $|\psi\rangle_{AB}$ in vector space $H_A\otimes H_B$, the concurrence is given by \cite{AU,PR,SA}
\begin{equation}\label{CD}
C(|\psi\rangle_{AB})=\sqrt{{2\left[1-\mathrm{Tr}(\rho_A^2)\right]}},
\end{equation}
where $\rho_A$ is the reduced density matrix by tracing over the subsystem $B$, $\rho_A=\mathrm{Tr}_B(|\psi\rangle_{AB}\langle\psi|)$. The concurrence for a bipartite mixed state $\rho_{AB}$ is defined by the convex roof
\begin{equation*}
 C(\rho_{AB})=\min_{\{p_i,|\psi_i\rangle\}}\sum_ip_iC(|\psi_i\rangle),
\end{equation*}
where the minimum is taken over all possible decompositions of $\rho_{AB}=\sum_ip_i|\psi_i\rangle\langle\psi_i|$, with $p_i\geq0$, $\sum_ip_i=1$ and $|\psi_i\rangle\in H_A\otimes H_B$.

For a tripartite state $|\psi\rangle_{ABC}$, the concurrence of assistance (COA) is defined by \cite{CH}
\begin{equation*}
 C_a(|\psi\rangle_{ABC})=C_a(\rho_{AB})=\max_{\{p_i,|\psi_i\rangle\}}\sum_ip_iC(|\psi_i\rangle),
\end{equation*}
for all possible ensemble realizations of $\rho_{AB}=\mathrm{Tr}_C(|\psi\rangle_{ABC}\langle \psi|)=\sum_ip_i|\psi_i\rangle\langle\psi_i|$. When $\rho_{AB}$ is a pure state, one has $C(|\psi\rangle_{AB})=C_a(\rho_{AB})$.

For an $N-$qubit state $|\psi\rangle_{AB_1,\cdots,B_{N-1}}\in H_A\otimes H_{B_1}\otimes\cdots\otimes H_{B_{N-1}}$, the concurrence $C(|\psi\rangle_{A|B_1\cdots B_{N-1}})$ of the state $|\psi\rangle_{A|B_1\cdots B_{N-1}}$, viewed as a bipartite with partitions $A$ and $B_1B_2\cdots B_{N-1}$, satisfies the monogamy inequality for $\alpha\geq2$ \cite{TF},
\begin{eqnarray}\label{CAA}
 &&C^{2}(\rho_{A|B_1,B_2\cdots,B_{N-1}})\nonumber\\&&\geq C^{2}(\rho_{AB_1})+C^{2}(\rho_{AB_2})+\cdots+C^{2}(\rho_{AB_{N-1}}),
\end{eqnarray}
where $C(\rho_{AB_i})$ is the concurrence of $\rho_{AB_i}=\mathrm{Tr}_{B_1\cdots B_{i-1}B_{i+1}\cdots B_{N-1}}(\rho)$.

The dual inequality satisfied by COA for $N$-qubit states has the form \cite{GSB},
\begin{eqnarray}\label{DCA}
 C^2(|\psi\rangle_{A|B_1B_2\cdots B_{N-1}}) \leq \sum_{i=1}^{N-1}C_a^2(\rho_{AB_i}).
\end{eqnarray}

Furthermore, the authors in \cite{XS} give an generalized monogamy relation for $\alpha\geq2$,
$C^{\alpha}(\rho_{A|B_1,B_2\cdots,B_{N-1}})\geq C^{\alpha}(\rho_{AB_1})+C^{\alpha}(\rho_{AB_2})+\cdots+C^{\alpha}(\rho_{AB_{N-1}})$. 
However, there are no dual inequalities yet satisfied by the $\alpha$th power of COA for $N$-qubit states. 
In this paper, we will give some monogamy and polygamy relations for $N$-qubit states in terms of the $\alpha$th power of COA.

The concurrence (\ref{CD}) is related to the linear entropy $T(\rho)$ of a state $\rho$, $T(\rho)=1-\mathrm{Tr}(\rho^2)$ \cite{EM}.
For a bipartite state $\rho_{AB}$, $T (\rho)$ has the property \cite{CYY}
\begin{eqnarray}\label{LS}
|T(\rho_A)-T(\rho_B)|\leq T(\rho_{AB})\leq T(\rho_A)+T(\rho_B).
\end{eqnarray}

For convenience, we write (\ref{DCA}) as follows
\begin{eqnarray}\label{coa}
 C^2(|\psi\rangle_{A|B_1B_2\cdots B_{N-1}}) \leq \sum_{i=1}^{k}C_a^2(\rho_{AM_i}),
\end{eqnarray}
where $C_a^2(\rho_{AM_i})=\sum_{j=M_{i-1}+1}^{M_i}C_a^2(\rho_{AB_j})$ with $M_0=0,~\sum_{i=1}^kM_i=N-1$, $1\leq k\leq N-1$.
The summation on the right hand side of (\ref{coa}) has been separated into $k$ parts.  There is always a choice of $M_i$, such that the above relations is true.

{\bf[Theorem 1]}. For any $N$-qubit pure state $|\psi\rangle_{AB_1B_2\cdots B_{N-1}}$, we have
\begin{eqnarray}\label{le}
  &&C^\alpha(|\psi\rangle_{A|B_1B_2\cdots B_{N-1}})\nonumber\\
 &&\leq  C^\alpha_a(\rho_{AM_1})+\frac{\alpha}{2}C_a^\alpha(\rho_{AM_2})+\cdots\nonumber\\
 &&
  +\left(\frac{\alpha}{2}\right)^{k-1}C_a^\alpha(\rho_{AM_k}),
\end{eqnarray}
for all $0\leq \alpha\leq2$.

{\sf[Proof]}. Without loss of generality, we can always assume that $C_a^2(\rho_{AM_t})\geq \sum_{l=t+1}^k C_a^2(\rho_{AM_l})$, $1\leq t\leq k-1,~2\leq k\leq N-1$,  by reordering 
$M_1,M_2,\cdots,M_k$ and/or relabeling the subsystems in need. 
Form the result in \cite{GSB}, we have
\begin{eqnarray}\label{pfth1}
 &&C^\alpha(|\psi\rangle_{A|B_1B_2\cdots B_{N-1}}) \nonumber\\
 &&\leq \left(C_a^2(\rho_{AM_1})+\sum_{i=2}^{k}C_a^2(\rho_{AM_i})\right)^{\frac{\alpha}{2}}\nonumber\\
 &&=C_a^\alpha(\rho_{AM_1})\left(1+\frac{\sum_{i=2}^{k}C_a^2(\rho_{AM_i})}{C_a^2(\rho_{AM_1})}\right)^{\frac{\alpha}{2}}\nonumber \\
   && \leq C_a^\alpha(\rho_{AM_1})\left[1+\frac{\alpha}{2}\left(\frac{\sum_{i=2}^{k}C_a^2(\rho_{AM_i})}{C_a^2(\rho_{AM_1})}\right)^{\frac{\alpha}{2}}\right]\nonumber\\
   &&=C^\alpha_a(\rho_{AM_1})+\frac{\alpha}{2}\left(\sum_{i=2}^{k}C_a^2(\rho_{AM_i})\right)^\frac{\alpha}{2}\nonumber\\
   &&\leq \cdots \leq\sum_{i=1}^k\left(\frac{\alpha}{2}\right)^{i-1}C^\alpha_a(\rho_{AM_i}),
\end{eqnarray}
where the first inequality is due to (\ref{coa}). Using the inequality $(1+t)^x\leq 1+xt \leq 1+xt^x$ for $0\leq x\leq1,~0\leq t\leq1$, we get the second inequality. $\Box$

Theorem 1 gives the polygamy relation of the $\alpha$th ($0\leq \alpha\leq2$) power of concurrence for $N$-qubit pure state $|\psi\rangle_{A|B_1B_2,\cdots,B_{N-1}}$ based on the COA. For the case of $k=N-1$, we have the following result,
\begin{eqnarray}\label{pfth1}
 &&C^\alpha(|\psi\rangle_{A|B_1B_2\cdots B_{N-1}})\nonumber\\&&
 \leq C^\alpha_a(\rho_{AB_1})+\frac{\alpha}{2}C_a^\alpha(\rho_{AB_2})+\cdots\nonumber\\&&
 +\left(\frac{\alpha}{2}\right)^{N-2}C_a^\alpha(\rho_{AB_{N-1}}).
\end{eqnarray}
Specially, for $\alpha=2$, (\ref{le}) or (\ref{pfth1}) reduces to the result (\ref{coa}) in \cite{GSB}.

{\it Example 1}. Let us consider the three-qubit state $|\psi\rangle$, which can be written in the generalized Schmidt decomposition form \cite{AA,XH},
\begin{equation*}
|\psi\rangle=\lambda_0|000\rangle+\lambda_1e^{i{\varphi}}|100\rangle+\lambda_2|101\rangle
+\lambda_3|110\rangle+\lambda_4|111\rangle,
\end{equation*}
where $\lambda_i\geq0,~i=0,\cdots,4$ and $\sum_{i=0}^4\lambda_i^2=1.$
We have $C(\rho_{A|BC})=2\lambda_0\sqrt{{\lambda_2^2+\lambda_3^2+\lambda_4^2}}$, $C(\rho_{AB})=2\lambda_0\lambda_2$, $C(\rho_{AC})=2\lambda_0\lambda_3$,  $C_a(\rho_{AB})=2\lambda_0\sqrt{{\lambda_2^2+\lambda_4^2}}$,  $C_a(\rho_{AC})=2\lambda_0\sqrt{{\lambda_3^2+\lambda_4^2}}$.
Set $\lambda_0=\lambda_1=\lambda_2=\lambda_3=\lambda_4=\frac{\sqrt{5}}{5}$. One gets $C^{\alpha}(\rho_{A|BC})=(\frac{2\sqrt{3}}{5})^{\alpha}$, $C_a^{\alpha}(\rho_{AB})+\frac{\alpha}{2}C_a^{\alpha}(\rho_{AC})=\left(1+\frac{\alpha}{2}\right)(\frac{2\sqrt{2}}{5})^{\alpha}$.
The relation from (\ref{le}) is shown in Fig. 1.

\begin{figure}
  \centering
  \includegraphics[width=7cm]{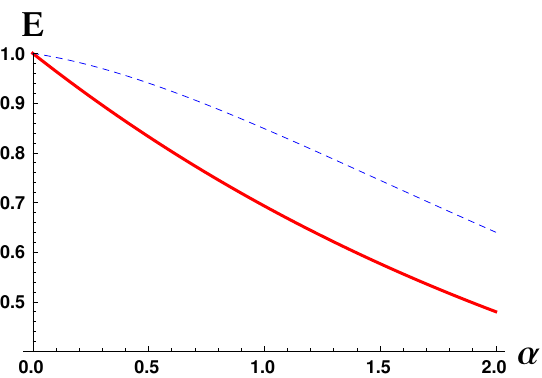}\\
  \caption{$E$ is the entanglement as a function of $\alpha$. Solid line is the $\alpha$th power of concurrence; Dashed line is the upper bound -- the right hand side of (\ref{le}).}\label{1}
\end{figure}

{\sf [Remark 1].} We have obtained a general polygamy inequality based on the $\alpha$th ($0\leq \alpha \leq 2$) power of concurrence, which includes the standard inequality (5) as a special case of  $\alpha=2$. 
Theorem 1 gives a weighted polygamy inequality, which can give rise to finer characterizations of entanglement distributions. Namely, once the information about the ordering of the concurrence of assistance
between $A$ and parts of the systems $B_1B_2\cdots B_{N-1}$ are given, we can present better (weighted) characterizations of entanglement distribution among the subsystems.
For example, let us consider a three-qubit state $\rho_{ABC}\in H_{ABC}$. By Theorem 1, one has $C^\alpha(\rho_{A|BC})\leq C_a^\alpha(\rho_{AB})+\alpha/2C_a^\alpha(\rho_{AC})$ 
if $C_a^\alpha(\rho_{AB})\geq C_a^\alpha(\rho_{AC})$. Since $\alpha/2\leq 1$, the weight of entanglement between subsystems $AB$ is larger than that between subsystems $AC$. 
For the case that $\alpha=1$, we have $C(\rho_{A|BC})\leq C_a(\rho_{AB})+1/2C_a(\rho_{AC})$. For given $C(\rho_{A|BC})$, if $C_a(\rho_{AB})$ reduces a mount $\Delta$, then $C_a(\rho_{AC})$
may need to increase $2\Delta$ to keep the inequality, which is not the case from (5), where the entanglements of subsystems $AB$ and $AC$ are equally weighted.

{\bf [Lemma 1]}. For arbitrary two real numbers $x,~y$ such that $x\geq y\geq0$, we have
 $(x-y)^\alpha\geq x^\alpha-y^\alpha$ and $(x+y)^\alpha\leq x^\alpha+y^\alpha$ for $0\leq\alpha\leq 1$.

{\sf [Proof]}. $(x-y)^\alpha\geq x^\alpha-y^\alpha$ is equivalent to $(1-\frac{y}{x})^\alpha+(\frac{y}{x})^\alpha\geq1$ for nonzero $x$. Denote $t=\frac{y}{x}$. Then $0\leq t\leq 1$. Set $f(t)=(1-t)^\alpha+t^\alpha$.
We have $\frac{\mathrm{d}f}{\mathrm{d}t}=\alpha[t^{\alpha-1}-(1-t)^{\alpha-1}]$.
When $0\leq t \leq \frac{1}{2}$, then $\frac{\mathrm{d}f}{\mathrm{d}t}\geq 0$, since $1-t\geq t$ and $\alpha-1<0$. Therefore, $f(t)\geq f(0)=1$ in this case. When $\frac{1}{2}< t \leq 1$, then $\frac{\mathrm{d}f}{\mathrm{d}t}\leq 0$, since $t\geq 1-t$ and $\alpha-1\leq 0$. Hence, $f(t)\geq f(1)=1$ in this case.
In summary, for $0<\alpha\leq 1$, $f(t)_\mathrm{min}\geq f(0)=f(1)=1$. Similarly, one can get the second inequality in Lemma 1. When $\alpha=0$ or $x=0$, the inequality is trivial. Hence we complete the proof of the Lemma 1.

In the following, by using the conclusion of Theorem 1 and Lemma 1, we present some monogamy inequalities and lower bounds of concurrence in terms of concurrence and COA. These monogamy relations are satisfied by the concurrence of $N$-qubit states under the partition $AB$ and $C_1\cdots C_{N-2}$, as well as under the partition $ABC_1$ and $C_2\cdots C_{N-2}$, which generalize the monogamy inequalities for pure states in \cite{ZXN}.

{\bf [Theorem 2]}. For any $2\otimes2\otimes\cdots\otimes2$ pure state $\rho_{ABC_1\cdots C_{N-2}}$, we have
\begin{eqnarray}\label{thm2}\nonumber
&&C^\alpha(\rho_{AB|C_1\cdots C_{N-2}})\\ \nonumber
&&\geq\mathrm{max}\left\{\left(\sum_{i=1}^{N-2}C^2(\rho_{AC_i})+C^2(\rho_{AB})\right)^\frac{\alpha}{2}-J_B,\right.\\
&& \left.\left(\sum_{i=1}^{N-2}C^2(\rho_{BC_i})+C^2(\rho_{AB})\right)^\frac{\alpha}{2}-J_A\right\},
\end{eqnarray}
for $0\leq\alpha\leq2$, $N\geq 4$,
where $C_a^2(\rho_{AM_i})$ is defined in (\ref{coa}) and  $J_A= \sum_{i=1}^{k_1}\left(\frac{\alpha}{2}\right)^{i-1}C^\alpha_a(\rho_{AM_i})$, $J_B=\sum_{i=1}^{k_2}\left(\frac{\alpha}{2}\right)^{i-1}C^\alpha_a(\rho_{BM_i})$.

{\sf [Proof]}. Without loss of generality, there always exists a proper ordering of the subsystems $M_{t_i},M_{t_i+1},\cdots,M_{k_i}$ $(i=1,2)$ such that $C_a^2(\rho_{AM_{t_1}})\geq \sum_{l={t_1}+1}^{k_1} C_a^2(\rho_{AM_l})$ and $C_a^2(\rho_{BM_{t_2}})\geq \sum_{l={t_2}+1}^{k_2} C_a^2(\rho_{BM_l})$, $1\leq t_1,t_2\leq k-1$, $2\leq k_1,k_2\leq N-1$.

For $2\otimes2\otimes\cdots\otimes2$ pure state $\rho_{ABC_1\cdots C_{N-2}}$, if $C^2(\rho_{A|BC_1\cdots C_{N-2}})\geq C^2(\rho_{B|AC_1\cdots C_{N-2}})$, one has
\begin{eqnarray*}
&&C^\alpha(\rho_{AB|C_1\cdots C_{N-2}})\\&&=(2T(\rho_{AB}))^{\frac{\alpha}{2}}\\
&&\geq|2T(\rho_A)-2T(\rho_{B})|^{\frac{\alpha}{2}}\\
&&=|C^2(\rho_{A|BC_1\cdots C_{N-2}})-C^2(\rho_{B|AC_1\cdots C_{N-2}})|^{\frac{\alpha}{2}}\\
&&\geq C^\alpha(\rho_{A|BC_1\cdots C_{N-2}})-C^\alpha(\rho_{B|AC_1\cdots C_{N-2}})\\
&&\geq \left(\sum_{i=1}^{N-2}C^2(\rho_{AC_i})+C^2(\rho_{AB})\right)^\frac{\alpha}{2}-C^\alpha(\rho_{B|AC_1\cdots C_{N-2}})\\
&&\geq \left(\sum_{i=1}^{N-2}C^2(\rho_{AC_i})+C^2(\rho_{AB})\right)^\frac{\alpha}{2}-J_B,
\end{eqnarray*}
where the first inequality is due to the left inequality in (\ref{LS}). Using the Lemma 1, one gets the second inequality. From (\ref{CAA}) we get the third inequality. The last inequality is obtained by Theorem 1.

If $C^2(\rho_{A|BC_1\cdots C_{N-2}})\leq C^2(\rho_{B|AC_1\cdots C_{N-2}})$, similar to the above derivation, we can obtain another inequality in Theorem 2. $\Box$

{\it Example 2}. Let us consider the following four-qubit pure state,
\begin{eqnarray}\label{FS}
|\psi\rangle_{ABCD}=\frac{1}{\sqrt{3}}(|0000\rangle+|0010\rangle+|1011\rangle).
\end{eqnarray}
From the Theorem 1 in \cite{ZXN} we get $C^2(|\psi\rangle_{AB|CD})\geq\frac{4}{9}$ for $\alpha=2$. From our Theorem 2, we have $C^\alpha(|\psi\rangle_{AB|CD})\geq(\frac{2}{3})^\alpha$ for any $\alpha$, see Fig. 2.

\begin{figure}
  \centering
  \includegraphics[width=7cm]{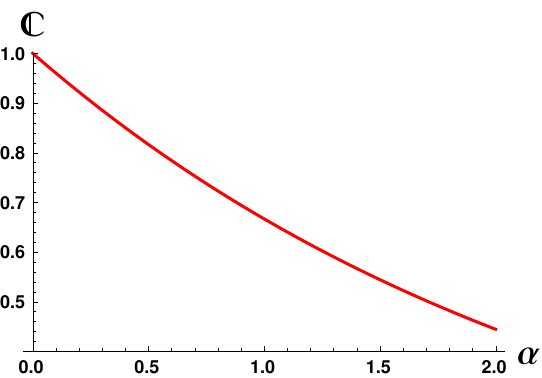}\\
  \caption{$C^\alpha(\rho_{AB|C_1\cdots C_{N-2}})$ as a function of $\alpha$ for (\ref{thm2}). The right end point of the curve corresponds to the value of $\alpha=2$, which is the result in \cite{ZXN}.}\label{2}
\end{figure}

{\sf [Remark 2].} Theorem 2 gives a monogamy inequality under the partition $AB$ and $C_1\cdots C_{N-2}$. Similar to Theorem 1, it also gives rise finer weighted characterizations of the entanglement distributions among the subsystems, as is illustrated in Example 2. Here we indicate that the proof of the main result in Ref. [25] has a flaw: where the equality $C^2(|\psi\rangle_{A|BC})=C^2(\rho_{AB})+C^2_a(\rho_{AC})$ is used 
for an arbitrary $2\otimes 2\otimes m$ quantum pure state $|\psi\rangle_{ABC}$, which has not been proved yet. We used some basic inequalities and Theorem 1 to prove the Theorem 2, which includes the result in Ref. [25]
as a special case of $\alpha=2$.

{\bf [Theorem 3]}. For any $2\otimes2\otimes\cdots\otimes2$ pure state $|\psi\rangle_{ABC_1\cdots C_{N-2}}$,  we have
\begin{eqnarray}\label{thm3}
C^\alpha(|\psi\rangle_{AB|C_1\cdots C_{N-2}})\leq J_A+J_B
\end{eqnarray}
for $0\leq\alpha\leq2$, $N\geq 4$, where $J_A$ and $J_B$ are defined the same with Theorem 2.

{\sf [Proof]}. Without loss of generality, there always exists a proper ordering of the subsystems $M_{t_i},M_{t_i+1},\cdots,M_{k_i}$ $(i=1,2)$ such that $C_a^2(\rho_{AM_{t_1}})\geq \sum_{l={t_1}+1}^{k_1} C_a^2(\rho_{AM_l})$ and $C_a^2(\rho_{BM_{t_2}})\geq \sum_{l={t_2}+1}^{k_2} C_a^2(\rho_{BM_l})$, $1\leq t_1,t_2\leq k-1$, $2\leq k_1,k_2\leq N-1$.

For $2\otimes2\otimes\cdots\otimes2$ pure state $|\psi\rangle_{ABC_1\cdots C_{N-2}}$, one has
\begin{eqnarray*}
&&C^\alpha(|\psi\rangle_{AB|C_1\cdots C_{N-2}})\\&&=(2T(\rho_{AB}))^{\frac{\alpha}{2}}\\
&&\leq(2T(\rho_A)+2T(\rho_{B}))^{\frac{\alpha}{2}}\\
&&=(C^2(\rho_{A|BC_1\cdots C_{N-2}})+C^2(\rho_{B|AC_1\cdots C_{N-2}}))^{\frac{\alpha}{2}}\\
&&\leq C^\alpha(\rho_{A|BC_1\cdots C_{N-2}})+C^\alpha(\rho_{B|AC_1\cdots C_{N-2}})\\
&&\leq J_A+J_B,
\end{eqnarray*}
where the first inequality is due to the right inequality in (\ref{LS}). The second inequality is due to Lemma 1. Using the Theorem 1, one gets the last inequality. $\Box$

For the four-qubit state (\ref{FS}), we have $C_a(\rho_{AB})=0$. Then from the Theorem 2 in \cite{ZXN}, we get $C^2(|\psi\rangle_{AB|CD})\leq\frac{4}{3}$ for $\alpha=2$. From our Theorem 3,
we have $C^\alpha(|\psi\rangle_{AB|CD})\leq (\frac{2\sqrt{2}}{3})^\alpha+\frac{\alpha}{2}(\frac{2}{3})^\alpha$ for any $\alpha$, see Fig. 3.

{\sf [Remark 3].} We have presented a weighted polygamy inequality for the $\alpha$th ($0\leq \alpha \leq 2$) power of concurrence, which includes the result in Ref. [25] as a special case of $\alpha=2$.
For $0\leq \alpha <2$, Theorem 3 gives rise to again finer characterizations of the entanglement distributions by weighting the entanglement of the subsystems.

\begin{figure}
  \centering
  \includegraphics[width=7cm]{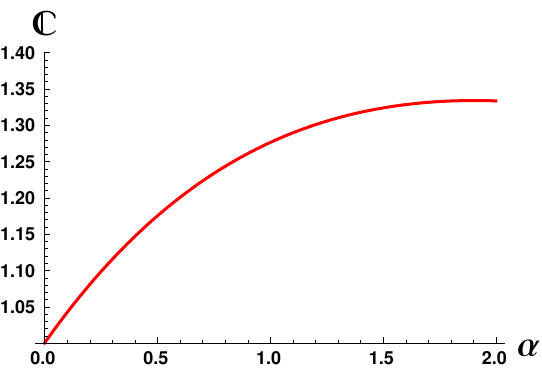}\\
  \caption{$C^\alpha(\rho_{AB|C_1\cdots C_{N-2}})$ as a function of $\alpha$ for (\ref{thm2}). The right end point of the curve corresponds to the value of $\alpha=2$, which is the result in \cite{ZXN}.}\label{3}
\end{figure}

Now we generalize our results to the concurrence $C_{ABC_1|C_2\cdots C_{N-2}}(|\psi\rangle)$ under the partition $ABC_1$ and $C_2 \cdots C_{N-2}~(N\geq6)$ for pure state $|\psi\rangle_{ABC_1\cdots C_{N-2}}$.
Similar to Theorem 2 and Theorem 3, we obtain the following corollaries:

{\bf [Corollary 1]}.  For any $N$-qubit pure state $|\psi\rangle_{ABC_1\cdots C_{N-2}}$, we have
\begin{eqnarray}\label{CA1}
&&C^\alpha(|\psi\rangle_{ABC_1|C_2\cdots C_{N-2}})\nonumber\\
&&\geq\mathrm{max}\left\{\left(\sum_{i=1}^{N-2}C^2(\rho_{AC_i})+C^2(\rho_{AB})\right)^\frac{\alpha}{2}-J_B,\right.\nonumber\\
&&\left. \left(\sum_{i=1}^{N-2}C^2(\rho_{BC_i})+C^2(\rho_{AB})\right)^\frac{\alpha}{2}-J_A\right\}-J_{C_1},
\end{eqnarray}
where $J_A,~J_B$ are defined as in Theorem 2, $J_{C_1}=\sum_{i=1}^k\left(\frac{\alpha}{2}\right)^{i-1}C^\alpha_a(\rho_{C_1M_i})$, $2\leq m\leq N-3,~N\geq6$.

{\sf [Proof]}. For any $N$-qubit pure state $|\psi\rangle_{ABC_1\cdots C_{N-2}}$, if $C^2(|\psi\rangle_{AB|C_1\cdots C_{N-2}})\geq C^2(|\psi\rangle_{C_1|ABC_2\cdots C_{N-2}})$, we have
\begin{eqnarray*}
&&C^\alpha(|\psi\rangle_{ABC_1|C_2\cdots C_{N-2}})\\&&=(2T(\rho_{ABC_1}))^{\frac{\alpha}{2}}\\
&&\geq|2T(\rho_{AB})-2T(\rho_{C_1})|^{\frac{\alpha}{2}}\\
&&=|C^2(|\psi\rangle_{AB|C_1\cdots C_{N-2}})- C^2(|\psi\rangle_{C_1|ABC_2\cdots C_{N-2}})|^{\frac{\alpha}{2}}\\
&&\geq C^\alpha(|\psi\rangle_{AB|C_1\cdots C_{N-2}})-C^\alpha(|\psi\rangle_{C_1|ABC_2\cdots C_{N-2}}),
\end{eqnarray*}
where the first inequality is due to $T(\rho_{ABC_1})\geq T(\rho_{AB})-T(\rho_{C_1})$. Using Lemma 1, we get the second inquality. Combining Theorem 1 and Theorem 2, we obtain (\ref{CA1}). $\Box$

{\bf [Corollary 2]}. For any $N$-qubit pure state $|\psi\rangle_{ABC_1\cdots C_{N-2}}$, if $C^2(|\psi\rangle_{AB|C_1\cdots C_{N-2}})\leq C^2(|\psi\rangle_{C_1|ABC_2\cdots C_{N-2}})$, we have
\begin{eqnarray}\label{CA2}
&&C^\alpha(|\psi\rangle_{ABC_1|C_2\cdots C_{N-2}})\nonumber\\
&&\geq\left(C^2(\rho_{AC_1})+C^2(\rho_{BC_1})+\sum_{i=2}^{N-2}C^2(\rho_{C_1C_i})\right)^{\frac{\alpha}{2}}\nonumber\\
&&-J_A-J_B,
\end{eqnarray}
and
\begin{eqnarray}\label{CA3}
C^\alpha(|\psi\rangle_{ABC_1|C_2\cdots C_{N-2}})\leq J_A+J_B+J_{C_1},
\end{eqnarray}
where $J_A,~J_B$ are defined in Theorem 2, $J_{C_1}$ is defined in Corollary 1.

In Corollary 2, the upper bound is due to the right inequalities of (\ref{LS}) and (\ref{le}). Analogously, by using $T(\rho_{ABC_1})\geq |T(\rho_{AC_1})-T(\rho_B)|,~T(\rho_{ABC_1})\geq |T(\rho_{A})-T(\rho_{BC_1})|$, and $T(\rho_{ABC_1})\leq |T(\rho_{AC_1})+T(\rho_B)|,~T(\rho_{ABC_1})\leq |T(\rho_{A})+T(\rho_{BC_1})|$, one can gets (\ref{CA2}) and (\ref{CA3}).

The lower bounds in Corollary 1 and Corollary 2 are not equivalent. We consider the following two examples to show that Corollary 1 and Corollary 2 give rise to different lower bounds.

{\it Example 3}. Let us consider the pure state $|\psi\rangle_{ABC_1C_2C_3C_4}=\frac{1}{\sqrt{2}}(|000000\rangle+|101000\rangle)$. We have $C(\rho_{AB})=C(\rho_{AC_i})=C_a(\rho_{AC_i})=C(\rho_{C_1C_i})=C_a(\rho_{C_1C_i})=0$ for $i=2,~3,~4$; $C(\rho_{BC_i})=C_a(\rho_{BC_i})=0$ for $i=1,~2,~3,~4$; and $C(\rho_{AC_1})=C_a(\rho_{AC_1})=1$. Thus, we have $C(|\psi\rangle)\geq1$ from (\ref{CA1}) and $C(|\psi\rangle)\geq0$ from (\ref{CA2}). Namely, bound (\ref{CA1}) is better than (\ref{CA2}) in this case.

{\it Example 4}. For the state $|\psi\rangle_{ABC_1C_2C_3C_4}=\frac{1}{\sqrt{2}}(|000000\rangle+|001100\rangle)$, it is straightforward to calculate that $C(\rho_{AB})=C(\rho_{AC_i})=C_a(\rho_{AC_i})=C_a(\rho_{BC_i})=C(\rho_{BC_i})=C(\rho_{C_1C_3})=C(\rho_{C_1C_4})=0$ for $i=2,~3,~4$; and $C(\rho_{C_1C_2})=C_a(\rho_{C_1C_2})=1$. Hence, we have $C(|\psi\rangle)\geq0$ from (\ref{CA1}) and $C(|\psi\rangle)\geq1$ from (\ref{CA2}). The bound (\ref{CA2}) is better than (\ref{CA1}) in this case.

\section{generalized monogamy and polygamy relations of negativity}

Given a bipartite state $\rho_{AB}$ in $H_A\otimes H_B$, the negativity is defined by \cite{GRF}
\begin{equation*}
 N(\rho_{AB})=\frac{||\rho_{AB}^{T_A}||-1}{2},
\end{equation*}
where $\rho_{AB}^{T_A}$ is the partially transposed matrix of $\rho_{AB}$ with respect to the subsystem $A$, $||X||$ denotes the trace norm of $X$, i.e $||X||=\mathrm{Tr}\sqrt{XX^\dag}$.
It vanishes if and only if $\rho_{AB}$ is separable for the $2\otimes2$ and $2\otimes3$ systems \cite{MPR}. We use the following definition of negativity for purposes of discussion:
\begin{equation*}
 N(\rho_{AB})=||\rho_{AB}^{T_A}||-1.
\end{equation*}
For any bipartite pure state $|\psi\rangle_{AB}$ in a $d\otimes d$ quantum system with Schmidt rank $d$,
$|\psi\rangle_{AB}=\sum_{i=1}^d\sqrt{\lambda_i}|ii\rangle$,
one has
\begin{eqnarray}\label{ne}
 N(|\psi\rangle_{AB})=2\sum_{i<j}\sqrt{\lambda_i\lambda_j},
\end{eqnarray}
from the definition of concurrence (\ref{CD}), we have
\begin{eqnarray}\label{co}
 C(|\psi\rangle_{AB})=2\sqrt{\sum_{i<j}\lambda_i\lambda_j}.
\end{eqnarray}
Combining (\ref{ne}) with (\ref{co}), one obtains
\begin{eqnarray}\label{nac}
N(|\psi\rangle_{AB})\geq C(|\psi\rangle_{AB}).
\end{eqnarray}

For any bipartite pure state    $|\psi\rangle_{AB}$ with Schmidt rank 2,
one has $N(|\psi\rangle_{AB})=C(|\psi\rangle_{AB})$ from (\ref{ne}) and (\ref{co}).
For a mixed state $\rho_{AB}$, the convex-roof extended negativity  (CREN) is defined by
\begin{equation*}
 N_c(\rho_{AB})=\mathrm{min}\sum_ip_iN(|\psi_i\rangle_{AB}),
\end{equation*}
where the minimum is taken over all possible pure state decompositions $\{p_i,~|\psi_i\rangle_{AB}\}$ of $\rho_{AB}$. CREN gives a perfect discrimination of positively partial transposed bound entangled states and separable states in any bipartite quantum systems \cite{PH,WJM}. For a mixed state $\rho_{AB}$, the convex-roof extended negativity of assistance (CRENOA) is defined by \cite{JAB}
\begin{equation*}
 N_a(\rho_{AB})=\mathrm{max}\sum_ip_iN(|\psi_i\rangle_{AB}),
\end{equation*}
where the maximum is taken over all possible pure state decompositions $\{p_i,~|\psi_i\rangle_{AB}\}$ of $\rho_{AB}$.

CREN is equivalent to concurrence for any pure state with Schmidt rank 2 \cite{JAB}. Consequently for any two-qubit mixed state $\rho_{AB}$, one has
\begin{eqnarray}\label{N1}
 N_c(\rho_{AB})=C(\rho_{AB})
\end{eqnarray}
and
\begin{eqnarray}\label{N2}
 N_a(\rho_{AB})= C_a(\rho_{AB}).
\end{eqnarray}

For $N$-qubit pure state $|\psi\rangle_{A|B_1B_2,\cdots,B_{N-1}}$, from (\ref{nac}), (\ref{N1}), (\ref{N2}) and the monogamy of the concurrence, we have
\begin{eqnarray}\label{negativity1}
&&N^\alpha(|\psi\rangle_{A|B_1B_2,\cdots,B_{N-1}})\nonumber\\&&\geq N_c^\alpha(\rho_{AB_1})+N_c^\alpha(\rho_{AB_2})+\cdots+N_c^\alpha(\rho_{AB_{N-1}}),
\end{eqnarray}
for $\alpha\geq2$.
The dual inequality \cite{JAB} in terms of CRENOA is given by
\begin{eqnarray}\label{negativity2}
&&N^2(|\psi\rangle_{A|B_1B_2,\cdots,B_{N-1}})\nonumber\\
&&\leq N_a^2(\rho_{AB_1})+N_a^2(\rho_{AB_2})+\cdots+N_a^2(\rho_{AB_{N-1}})\nonumber\\
&&=\sum_{i=1}^{k}N_a^2(\rho_{AM_i}),
\end{eqnarray}
where $N_a^2(\rho_{AM_i})=\sum_{j=M_{i-1}+1}^{M_i}N_a^2(\rho_{AB_j})$ with $M_0=0,~\sum_{i=1}^kM_i=N-1$, $1\leq k\leq N-1$. By similar consideration to concurrence, we get the upper bound of the $\alpha$th power of negativity as follows.

{\bf[Theorem 4]}. For any $N$-qubit pure state $|\psi\rangle_{AB_1B_2,\cdots,B_{N-1}}$, we have
\begin{eqnarray}\label{th4}
 && N^\alpha(|\psi\rangle_{A|B_1B_2,\cdots,B_{N-1}})\nonumber\\
  &&\leq  N^\alpha_a(\rho_{AM_1})+\frac{\alpha}{2}N_a^\alpha(\rho_{AM_2})+\cdots\nonumber\\
  &&+\left(\frac{\alpha}{2}\right)^{k-1}N_a^\alpha(\rho_{AM_k}),
\end{eqnarray}
for all $0\leq \alpha\leq2$, $N\geq 4$.

{\bf [Theorem 5]}. For any $2\otimes2\otimes\cdots\otimes2$ pure state $|\psi\rangle_{ABC_1\cdots C_{N-2}}$, we have
\begin{eqnarray*}
&&N^\alpha(|\psi\rangle_{AB|C_1\cdots C_{N-2}})\\
&&\geq\mathrm{max}(\left(\sum_{i=1}^{N-2}N_c^2(\rho_{AC_i})+N_c^2(\rho_{AB})\right)^\frac{\alpha}{2}-J'_B,\\&& \left(\sum_{i=1}^{N-2}N_c^2(\rho_{BC_i})+N_c^2(\rho_{AB})\right)^\frac{\alpha}{2}-J'_A),
\end{eqnarray*}
for $0<\alpha\leq2$, $N\geq 4$,
where $J'_A= \sum_{i=1}^{k_1}\left(\frac{\alpha}{2}\right)^{i-1}N^\alpha_a(\rho_{AM_i})$, $J'_B=\sum_{i=1}^{k_2}\left(\frac{\alpha}{2}\right)^{i-1}N^\alpha_a(\rho_{BM_i})$.

{\sf [Proof]}. Without loss of generality, there always exists a proper ordering of the subsystems $M_{t_i},M_{t_i+1},\cdots,M_{k_i}$ $(i=1,2)$ such that $N_a^2(\rho_{AM_{t_1}})\geq \sum_{l={t_1}+1}^{k_1} N_a^2(\rho_{AM_l})$ and $N_a^2(\rho_{BM_{t_2}})\geq \sum_{l={t_2}+1}^{k_2} N_a^2(\rho_{BM_l})$, $1\leq t_1,t_2\leq k-1$, $2\leq k_1,k_2\leq N-1$.

For $2\otimes2\otimes\cdots\otimes2$ pure state $|\psi\rangle_{ABC_1\cdots C_{N-2}}$, we have
\begin{eqnarray*}
&&N^\alpha(|\psi\rangle_{AB|C_1\cdots C_{N-2}})\geq C^\alpha(|\psi\rangle_{AB|C_1\cdots C_{N-2}})\nonumber\\
&&\geq\mathrm{max}(\left(\sum_{i=1}^{N-2}C^2(\rho_{AC_i})+C^2(\rho_{AB})\right)^\frac{\alpha}{2}-J_B,\\&& \left(\sum_{i=1}^{N-2}C^2(\rho_{BC_i})+C^2(\rho_{AB})\right)^\frac{\alpha}{2}-J_A)\nonumber\\
&&=\mathrm{max}(\left(\sum_{i=1}^{N-2}N_c^2(\rho_{AC_i})+N_c^2(\rho_{AB})\right)^\frac{\alpha}{2}-J'_B,\\&& \left(\sum_{i=1}^{N-2}N_c^2(\rho_{BC_i})+N_c^2(\rho_{AB})\right)^\frac{\alpha}{2}-J'_A),
\end{eqnarray*}
where the first inequality is due to (\ref{nac}), the second inequality is from Theorem 2, the equality is based on (\ref{N1}) and  (\ref{N2}). $\Box$

For $N$-qubit pure state $|\psi\rangle_{A|B_1B_2,\cdots,B_{N-1}}$, based on the result in \cite{ce,zzj}, $N(|\psi\rangle_{AB|C_1\cdots C_{N-2}})\leq \sqrt{\frac{r(r-1)}{2}}C(|\psi\rangle_{AB|C_1\cdots C_{N-2}})$, where $r$ is the Schmidt rank of the pure state $|\psi\rangle_{ABC_1\cdots C_{N-2}}$. From Theorem 3, we can obtain the upper bound of negativity under the partition $AB$ and $C_1\cdots C_{N-2}$.

{\bf [Theorem 6]}. For any $2\otimes2\otimes\cdots\otimes2$ pure state $|\psi\rangle_{ABC_1\cdots C_{N-2}}$, we have
\begin{eqnarray*}
N^\alpha(|\psi\rangle_{AB|C_1\cdots C_{N-2}})\leq \left(\frac{r(r-1)}{2}\right)^\frac{\alpha}{2}(J'_A+J'_B),
\end{eqnarray*}
for $0\leq\alpha\leq2$, where $J'_A,~J'_B$ are given in Theorem 5.

{\it Example 5}. Let us consider the 4-qubit generalized $W$-class state,
\begin{eqnarray}\nonumber
|W\rangle_{ABC_1C_2}&=\lambda_1|1000\rangle+\lambda_2|0100\rangle\\
&+\lambda_3|0010\rangle+\lambda_4|0001\rangle,
\end{eqnarray}
where $\sum_{i}\lambda_i^2=1$. We have $N(|W\rangle_{AB|C_1C_2})=2\sqrt{(\lambda_1^2+\lambda_2^2)(\lambda_3^2+\lambda_4^2)}$, $N_c(\rho_{AB})=N_a(\rho_{AB})=2\lambda_1\lambda_2$, $N_c(\rho_{AC_1})=N_a(\rho_{AC_1})=2\lambda_1\lambda_3$, $N_c(\rho_{AC_2})=N_a(\rho_{AC_2})=2\lambda_1\lambda_4$. Taking $\lambda_1=\frac{3}{4},~\lambda_2=\frac{1}{2},~\lambda_3=\frac{\sqrt{2}}{4},~\lambda_4=\frac{1}{4}$,
we get $J_A'=J_B'=\left(\frac{3}{4}\right)^\alpha+\frac{\alpha}{2}\left(\frac{3\sqrt{2}}{8}\right)^\alpha+\left(\frac{\alpha}{2}\right)^2\left(\frac{3}{8}\right)^\alpha$. Set $y=N^\alpha(|W\rangle_{AB|C_1C_2})-\left((N_c^2(\rho_{AB})+N_c^2(\rho_{AC_1})+N_c^2(\rho_{AC_2}))^\frac{\alpha}{2}-J_A'\right)$, i.e., the difference between the left and right side of Theorem 5.
We elucidate the results of Theorem 5 in Fig. 4 for this case. For the case that the Schmidt rank of the pure state of $|\psi\rangle_{AB|C_1\cdots C_{N-2}}$ is 2, we show the results of Theorem 6 in Fig. 5.
\begin{figure}
  \centering
  \includegraphics[width=7cm]{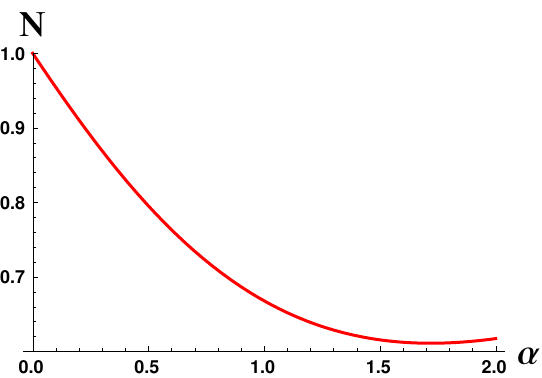}\\
  \caption{$N$ is the difference between the left and right side of Theorem 5}\label{3}
\end{figure}
\begin{figure}
  \centering
  \includegraphics[width=7cm]{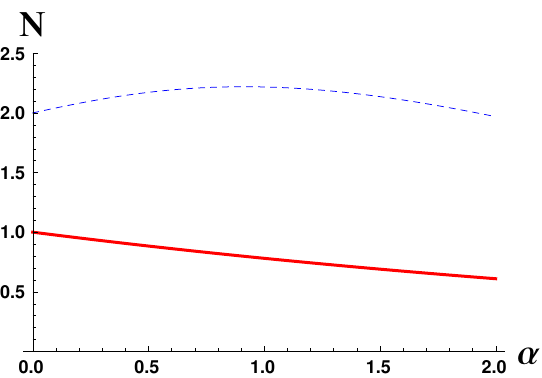}\\
  \caption{$N$ is the entanglement as a function of $\alpha$. Solid line is the value of $N^\alpha(|W\rangle_{AB|C_1C_2})$; Dashed line is the value of $J_A'+J_B'$.}\label{3}
\end{figure}

\section{conclusion}
Entanglement monogamy and polygamy are fundamental properties of multipartite entangled states. We have presented the monogamy relations of the $\alpha$th power of concurrence for $N$-qubit systems by showing the relations among $C(|\psi\rangle_{AB|C_1\cdots C_{N-2}}),~C(\rho_{AB}),~C(\rho_{AC_i}),~C(\rho_{BC_i}),~C_a(\rho_{AC_i})$, and $C_a(\rho_{BC_i})$, $1\leq i\leq N-2$, which give rise to the lower and upper bounds on the entanglement sharing among the partitions. We have investigated the monogamy relations based on concurrence and COA. The upper bound of the $\alpha$th ($0\leq\alpha\leq2$) power of concurrence has been obtained based on the COA.
We have obtained the monogamy and polygamy relations satisfied by the $\alpha$th ($0\leq\alpha\leq2$) power of concurrence in $N$-qubit pure states under the partition $AB$ and $C_1 \cdots C_{N-2}$,
as well as under the partition $ABC_1$ and $C_1\cdots C_{N-2}$. These relations also give rise to a kind of trade-off relations related to the lower and upper bounds of concurrences.
Based on the relations between negativity and concurrence, we have also obtained the similar results for CREN and CRENOA.
These results may be generalized to monogamy and polygamy relations under arbitrary partitions $C_{ABC_1\cdots C_i|C_{i+1}\cdots C_{N-2}},~2\leq i\leq N-2$.
Our approach may be used for the investigation of entanglement distribution based on other measures of quantum correlations.

\bigskip
\noindent{\bf Acknowledgments}\, \, This work is supported by the NSF of China under Grant No. 11675113 and NSF of Beijing under No. KZ201810028042.

\end{document}